



\font\twelverm=cmr10 scaled 1200    \font\twelvei=cmmi10 scaled 1200
\font\twelvesy=cmsy10 scaled 1200   \font\twelveex=cmex10 scaled 1200
\font\twelvebf=cmbx10 scaled 1200   \font\twelvesl=cmsl10 scaled 1200
\font\twelvett=cmtt10 scaled 1200   \font\twelveit=cmti10 scaled 1200
\font\twelvesc=cmcsc10 scaled 1200

\skewchar\twelvei='177   \skewchar\twelvesy='60


\def\twelvepoint{\normalbaselineskip=12.4pt
  \abovedisplayskip 12.4pt plus 3pt minus 6pt
  \belowdisplayskip 12.4pt plus 3pt minus 6pt
  \abovedisplayshortskip 0pt plus 3pt
  \belowdisplayshortskip 7.2pt plus 3pt minus 4pt
  \smallskipamount=3.6pt plus1.2pt minus1.2pt
  \medskipamount=7.2pt plus2.4pt minus2.4pt
  \bigskipamount=14.4pt plus4.8pt minus4.8pt
  \def\rm{\fam0\twelverm}          \def\it{\fam\itfam\twelveit}%
  \def\sl{\fam\slfam\twelvesl}     \def\bf{\fam\bffam\twelvebf}%
  \def\mit{\fam 1}                 \def\cal{\fam 2}%
  \def\tt{\twelvett}
  \def\sc{\twelvesc}
  \def\nullspace{\nulldelimiterspace=0pt \mathsurround=0pt }
  \def\big##1{{\hbox{$\left##1\vbox to 10.2pt{}\right.\nullspace$}}}
  \def\Big##1{{\hbox{$\left##1\vbox to 13.8pt{}\right.\nullspace$}}}
  \def\bigg##1{{\hbox{$\left##1\vbox to 17.4pt{}\right.\nullspace$}}}
  \def\Bigg##1{{\hbox{$\left##1\vbox to 21.0pt{}\right.\nullspace$}}}
  \textfont0=\twelverm   \scriptfont0=\tenrm   \scriptscriptfont0=\sevenrm
  \textfont1=\twelvei    \scriptfont1=\teni    \scriptscriptfont1=\seveni
  \textfont2=\twelvesy   \scriptfont2=\tensy   \scriptscriptfont2=\sevensy
  \textfont3=\twelveex   \scriptfont3=\twelveex  \scriptscriptfont3=\twelveex
  \textfont\itfam=\twelveit
  \textfont\slfam=\twelvesl
  \textfont\bffam=\twelvebf \scriptfont\bffam=\tenbf
  \scriptscriptfont\bffam=\sevenbf
  \normalbaselines\rm}



\def\beginlinemode{\endmode
  \begingroup\parskip=0pt \obeylines\def\\{\par}\def\endmode{\par\endgroup}}
\def\beginparmode{\endmode
  \begingroup \def\endmode{\par\endgroup}}
\let\endmode=\par
{\obeylines\gdef\
{}}
\def\singlespace{\baselineskip=\normalbaselineskip}

\def\oneandahalfspace{\baselineskip=\normalbaselineskip
  \multiply\baselineskip by 3 \divide\baselineskip by 2}
\def\doublespace{\baselineskip=\normalbaselineskip \multiply\baselineskip by 2}

\newcount\firstpageno
\firstpageno=2
\footline={\ifnum\pageno<\firstpageno{\hfil}\else{\hfil\twelverm\folio\hfil}\fi}
\let\rawfootnote=\footnote		
\def\footnote#1#2{{\rm\singlespace\hang
  \rawfootnote{#1}{#2\hfill\vrule height 0pt depth 6pt width 0pt}}}

\def\raggedcenter{\leftskip=4em plus 12em \rightskip=\leftskip
  \parindent=0pt \parfillskip=0pt \spaceskip=.3333em \xspaceskip=.5em
  \pretolerance=9999 \tolerance=9999
  \hyphenpenalty=9999 \exhyphenpenalty=9999 }
\def\dateline{\rightline{\ifcase\month\or
  January\or February\or March\or April\or May\or June\or
  July\or August\or September\or October\or November\or December\fi
  \space\number\year}}
\def\received{\vskip 3pt plus 0.2fill
 \centerline{\sl (Received\space\ifcase\month\or
  January\or February\or March\or April\or May\or June\or
  July\or August\or September\or October\or November\or December\fi
  \qquad, \number\year)}}


\hsize=6.5truein
\vsize=8.9truein
\parskip=\medskipamount
\twelvepoint		
\doublespace		
\overfullrule=0pt	


\def\preprintno#1{
 \rightline{\rm #1}}	

\def\title			
  {\null\vskip 3pt plus 0.3fill
   \beginlinemode \doublespace \raggedcenter \bf}

\def\author			
  {\vskip 3pt plus 0.3fill \beginlinemode
   \oneandahalfspace \raggedcenter\sc}

\def\and			
  {\vskip 3pt plus 0.3fill \beginlinemode
   \oneandahalfspace \raggedcenter \rm and}

\def\affil			
  {\vskip 3pt plus 0.1fill \beginlinemode
   \oneandahalfspace \raggedcenter \sl}

\def\abstract			
  {\vskip 3pt plus 0.3fill \beginparmode
   \oneandahalfspace \narrower ABSTRACT:~~}

\def\endtitlepage		
  {\endpage			
   \body}

\def\body			
  {\beginparmode}		

\def\head#1{			
  \filbreak\vskip 0.5truein	
  {\immediate\write16{#1}
   \raggedcenter \uppercase{#1}\par}
   \nobreak\vskip 0.25truein\nobreak}

\def\subhead#1{			
  \vskip 0.25truein		
  {\raggedcenter #1 \par}
   \nobreak\vskip 0.25truein\nobreak}



\def\references
  {\subhead{REFERENCES}
   \frenchspacing \parindent=0pt \leftskip=0.8truecm \rightskip=0truecm
   \parskip=4pt plus 2pt \everypar{\hangindent=\parindent}}

\def\refstylenp{		
  \gdef\refto##1{~[##1]}				
  \gdef\r##1{~[##1]}	         			
  \gdef\refis##1{\indent\hbox to 0pt{\hss[##1]~}}     	
  \gdef\citerange##1##2##3{~[\cite{##1}--\setbox0=\hbox{\cite{##2}}\cite{##3}]}
  \gdef\journal##1, ##2, ##3,                           
    ##4{{##1} {##2} (##3) ##4}}

\def\refstylepr{		
  \gdef\refto##1{$^{##1}$}		
  \gdef\r##1{$^{##1}$}		        
  \gdef\refis##1{\indent\hbox to 0pt{\hss##1.~}}	
  \gdef\citerange##1##2##3{$^\cite{##1}-\setbox0=\hbox{\cite{##2}}\cite{##3}$}
  \gdef\journal##1, ##2, ##3,           		
    ##4.{{##1} {##2}, ##4 (##3).}}

\def\prd{\journal Phys. Rev. D}

\def\prl{\journal Phys. Rev. Lett.}

\def\np{\journal Nucl. Phys.}

\def\pl{\journal Phys. Lett.}

\def\prep{\journal Phys. Rep.}

\def\apj{\journal Astrophys. J.}

\def\endreferences{\body}

\def\figurecaptions		
  {\endpage
   \beginparmode
   \head{Figure Captions}
}

\def\endpage			
  {\vfill\eject}

\def\endpaper			
  {\endmode\vfill\supereject}


\def\ref#1{Ref. #1}			
\def\Ref#1{Ref. #1}			

\def\frac#1#2{{\textstyle{#1 \over #2}}}

\def\sla{\raise.15ex\hbox{$/$}\kern-.57em}
\def\leaderfill{\leaders\hbox to 1em{\hss.\hss}\hfill}
\def\twiddle{\lower.9ex\rlap{$\kern-.1em\scriptstyle\sim$}}
\def\bigtwiddle{\lower1.ex\rlap{$\sim$}}
\def\gtwid{\mathrel{\raise.3ex\hbox{$>$\kern-.75em\lower1ex\hbox{$\sim$}}}}
\def\ltwid{\mathrel{\raise.3ex\hbox{$<$\kern-.75em\lower1ex\hbox{$\sim$}}}}
\def\square{\kern1pt\vbox{\hrule height 1.2pt\hbox{\vrule width 1.2pt\hskip 3pt
   \vbox{\vskip 6pt}\hskip 3pt\vrule width 0.6pt}\hrule height 0.6pt}\kern1pt}
\def\ucsb{Department of Physics\\University of California\\
Santa Barbara, CA 93106}

\def\tev{{\rm \,Te\kern-0.125em V}}
\def\gev{{\rm \,Ge\kern-0.125em V}}
\def\mev{{\rm \,Me\kern-0.125em V}}
\def\kev{{\rm \,ke\kern-0.125em V}}
\def\ev{{\rm \,e\kern-0.125em V}}

\refstylenp


\catcode`@=11
\newcount\r@fcount \r@fcount=0
\newcount\r@fcurr
\immediate\newwrite\reffile
\newif\ifr@ffile\r@ffilefalse
\def\w@rnwrite#1{\ifr@ffile\immediate\write\reffile{#1}\fi\message{#1}}

\def\writer@f#1>>{}
\def\referencefile{
  \r@ffiletrue\immediate\openout\reffile=\jobname.ref%
  \def\writer@f##1>>{\ifr@ffile\immediate\write\reffile%
    {\noexpand\refis{##1} = \csname r@fnum##1\endcsname = %
     \expandafter\expandafter\expandafter\strip@t\expandafter%
     \meaning\csname r@ftext\csname r@fnum##1\endcsname\endcsname}\fi}%
  \def\strip@t##1>>{}}

\def\citeall#1{\xdef#1##1{#1{\noexpand\cite{##1}}}}
\def\cite#1{\each@rg\citer@nge{#1}}	

\def\each@rg#1#2{{\let\thecsname=#1\expandafter\first@rg#2,\end,}}
\def\first@rg#1,{\thecsname{#1}\apply@rg}	
\def\apply@rg#1,{\ifx\end#1\let\next=\relax
\else,\thecsname{#1}\let\next=\apply@rg\fi\next}

\def\citer@nge#1{\citedor@nge#1-\end-}	
\def\citer@ngeat#1\end-{#1}
\def\citedor@nge#1-#2-{\ifx\end#2\r@featspace#1 
  \else\citel@@p{#1}{#2}\citer@ngeat\fi}	
\def\citel@@p#1#2{\ifnum#1>#2{\errmessage{Reference range #1-#2\space is bad.}%
    \errhelp{If you cite a series of references by the notation M-N, then M and
    N must be integers, and N must be greater than or equal to M.}}\else%
 {\count0=#1\count1=#2\advance\count1
by1\relax\expandafter\r@fcite\the\count0,%
  \loop\advance\count0 by1\relax
    \ifnum\count0<\count1,\expandafter\r@fcite\the\count0,%
  \repeat}\fi}

\def\r@featspace#1#2 {\r@fcite#1#2,}	
\def\r@fcite#1,{\ifuncit@d{#1}
    \newr@f{#1}%
    \expandafter\gdef\csname r@ftext\number\r@fcount\endcsname%
                     {\message{Reference #1 to be supplied.}%
                      \writer@f#1>>#1 to be supplied.\par}%
 \fi%
 \csname r@fnum#1\endcsname}
\def\ifuncit@d#1{\expandafter\ifx\csname r@fnum#1\endcsname\relax}%
\def\newr@f#1{\global\advance\r@fcount by1%
    \expandafter\xdef\csname r@fnum#1\endcsname{\number\r@fcount}}

\let\r@fis=\refis			
\def\refis#1#2#3\par{\ifuncit@d{#1}
   \newr@f{#1}%
   \w@rnwrite{Reference #1=\number\r@fcount\space is not cited up to now.}\fi%
  \expandafter\gdef\csname r@ftext\csname r@fnum#1\endcsname\endcsname%
  {\writer@f#1>>#2#3\par}}

\def\ignoreuncited{
   \def\refis##1##2##3\par{\ifuncit@d{##1}%
     \else\expandafter\gdef\csname r@ftext\csname
r@fnum##1\endcsname\endcsname%
     {\writer@f##1>>##2##3\par}\fi}}

\def\r@ferr{\endreferences\errmessage{I was expecting to see
\noexpand\endreferences before now;  I have inserted it here.}}
\let\r@ferences=\references
\def\references{\r@ferences\def\endmode{\r@ferr\par\endgroup}}

\let\endr@ferences=\endreferences
\def\endreferences{\r@fcurr=0
  {\loop\ifnum\r@fcurr<\r@fcount
    \advance\r@fcurr by 1\relax\expandafter\r@fis\expandafter{\number\r@fcurr}%
    \csname r@ftext\number\r@fcurr\endcsname%
  \repeat}\gdef\r@ferr{}\endr@ferences}


\let\r@fend=\endpaper\gdef\endpaper{\ifr@ffile
\immediate\write16{Cross References written on []\jobname.REF.}\fi\r@fend}

\catcode`@=12

\citeall\refto		
\citeall\ref		%
\citeall\Ref		%


\ignoreuncited

\def\L{\Lambda}

\def\r{\refto}

\def\gev{{\rm \,Ge\kern-0.125em V}}
\def\mev{{\rm \,Me\kern-0.125em V}}
\def\prep{\journal Phys. Rep.}
\def\apj{\journal Astrophys. J.}

\singlespace

\preprintno{UCSBTH--94--11}
\preprintno{hep-ph/9407322}

\doublespace

\title
COSMOLOGICAL CONSTRAINTS ON THE SCALE OF SUPERSYMMETRY BREAKING

\author Raghavan Rangarajan$^{*}$
\affil\ucsb

\abstract
We consider the cosmological and astrophysical constraints on the
decay of massive $E_8 \times E_8^\prime$ superstring axions
associated with the hidden sector.  We find that decay lifetimes
greater than
1 s are ruled out by limits from
nucleosynthesis, by limits on the distortion of the cosmic microwave and
gamma ray backgrounds and by closure arguments.  We conclude that
$\Lambda>1.2\times10^{13}\gev$, where $\Lambda$ is the scale of gaugino
condensation
in the hidden sector.
This implies that the scale of supersymmetry breaking is greater than
$10^{10}\gev$.  Significantly, our result agrees with the
value of $5\times 10^{13}\gev$ for $\L$ obtained independently by
setting supersymmetric scalar masses equal to $m_W$.

(*: raghu@tpau.physics.ucsb.edu)
\vskip 1in
{\centerline {\it Nuclear Physics B 454 (1995) 357}}
\endtitlepage

\body

\baselineskip=17.5pt

$E_8 \times E^\prime_8$ superstring theories are anamoly free and hence very
attractive\r{ghmr,greenschwartz}.  Witten has
shown that in such theories there is a model
independent $U(1)$ Peccei-Quinn symmetry.  When this symmetry
breaks one gets an
axion with a decay constant
related
to the
compactification scale because the non-renormalizable interactions of the
model independent axion arise as a result of compactification\r{witten84}.
Furthermore, there are several model dependent axion degrees of freedom.

In $E_8 \times E^\prime_8$ models compactified on a
Calabi-Yau manifold
the model independent axion degree of freedom corresponds to $B_{\mu \nu}$,
$(\mu,\nu=1,..,4)$, the zero mode of the antisymmetric tensor field
$B_{MN}$ $(M,N=1,..,10)$ which is essential for anamoly
cancellation\r{witten84}.
The model dependent axion is a linear combination of the zero modes
$B_{mn}$ $(m,n=5,..,10)$\r{witten85b}.
Neither of the corresponding PQ symmetries survive compactification and
the above axions have decay
constants related to the compactification scale\r{choikim85a}.

If $ E^\prime_8$ breaks down to a non-abelian group there are two
non-abelian groups today: $SU(3)_C$ and the non-abelian subgroup of
$ E^\prime_8$.  The model independent and the model dependent axion degrees
of freedom rearrange themselves to give two physical axions-- the QCD axion
$a$ and the $ E^\prime_8$ axion $a^\prime$, with decay constants close
to the compactification scale\r{choikim85b}.
The energy density of the axions is proportional to $\Lambda^\alpha F_a^\beta$
$(\alpha, \beta >0)$,
where $\Lambda$ is the scale at which the underlying group gets strong.
The QCD axion with $F_a> 10^{12}\gev$ and $\Lambda\sim 200\mev$
will not have decayed by today and suffers an energy density problem.
In this paper we study the cosmological and astrophysical constraints on the
energy density of the $ E^\prime_8$ axion and on its decay to photons to
give limits on $\Lambda$ for the $ E^\prime_8$ axion.
We find that
$\Lambda$ should be greater than $1.2\times10^{13}\gev$.
$\Lambda$ is also the scale
of gaugino condensation
in the hidden sector which breaks supersymmetry\r{deribnil}.
Since the
masses of supersymmetric particles in the observed sector are approximately
$10^{-1}\Lambda^3/M_{Pl}^2$, this corresponds to supersymmetric particle masses
greater than $1\gev$.  The scale of supersymmetry breaking is given by
$M_S\sim (\Lambda^3/M_{Pl})^{1/2}$.  Thus, $M_S$ should be greater than
$10^{10}\gev$.
Our result agrees with
$\Lambda\simeq 5\times 10^{13}\gev$
obtained
by setting
$m_{\rm{gaugino}}\simeq m_W$\r{deribnil}.  We find it significant that
the result obtained by us using cosmological and astrophysical
constraints agrees with the independent requirement from particle
physics.

In a companion paper\r{ranga94a}, we set $\L$ equal to $5\times 10^{13}\gev$
and
study the dilution of the baryon asymmetry due to axion decays.  We find
that for this value of $\L$, most models of baryogenesis can not
tolerate the large dilution (by a factor of $10^7$) of the baryon asymmetry.

World sheet instanton effects may add terms to
the potential of the model dependent axion\refto{wenwit,dsww} for most
Calabi-Yau spaces (though not all).  In such a case, the QCD axion will
not get its mass purely from an $F\tilde{F}$ term but can have large
contributions from other non-derivative couplings.  The $E_8^\prime$
axion will be almost equivalent to $a_1$ below\r{kim87}.  The mass of the
$E_8^\prime$ axion will probably not be greatly affected.  Hence, we
ignore these effects below.

{\centerline{\bf II}}

In an $E_8\times E_8^\prime$ superstring theory compactified on a
Calabi-Yau manifold and in which the $E_8^\prime$ breaks down to a
non-abelian group, the axion lagrangian in 4 dimensions is
given by\r{choikim85b}
$$
\eqalignno{
\quad\quad\quad\quad\quad\quad\quad
L={1\over2}(\partial_\mu a_1)^2 +&{1\over2}(\partial_\mu a_2)^2 +
{1\over 32 \pi^2 F_1}a_1(g^2 F \tilde{F}
+ g^{\prime 2} F^\prime {\tilde{F}}^\prime)\cr
&+{1\over 32 \pi^2 F_2}a_2(g^2 F \tilde{F}
-g^{\prime 2} F^\prime {\tilde{F}}^\prime)
\quad\quad
\quad\quad\quad\quad\quad\quad\quad\quad (1)\cr}$$
where $a_1$ and $a_2$ are the
model independent and model dependent
axion degrees of freedom and $F_1$ and $F_2$
are the corresponding decay constants.  Both $F_1$ and $F_2$ are related
to the
compactification scale and are about $10^{15} \gev$\r{kim87}.
$F$ can be taken to be either the electromagnetic or weak or strong
interactions
field strength
tensor.
$F^\prime$ corresponds to hidden sector gauge field strength tensors.
The lagrangian can be rewritten as
$$
\eqalignno{\quad\quad
\quad\quad\quad\quad\quad
L={1\over2}(\partial_\mu a)^2 &+{1\over2}(\partial_\mu a^\prime)^2 +
{1\over 32 \pi^2 F_a}a(g^2 F \tilde{F} )\cr
&+\Bigl({1\over 32 \pi^2 F_a^\prime}a^\prime\Bigr)
\Bigl( g^{\prime 2} F^\prime {\tilde{F}}^\prime +
g^2 {F_2^2-F_1^2\over F_1^2+F_2^2} F \tilde F \Bigr)
\quad\quad\quad\quad
\quad\quad(2a)\cr}$$
where$$ a={F_1 a_1 + F_2 a_2 \over (F_1^2 + F_2^2)^{1/2}}, \eqno(2b)$$
    $$ a^\prime={F_2 a_1 - F_1 a_2 \over (F_1^2 + F_2^2)^{1/2}}, \eqno(2c)$$
      $$F_a={1\over 2}(F_1^2 + F_2^2)^{1/2}, \eqno(2d)$$
and  $$F_a^\prime=F_1 F_2/(F_1^2 + F_2^2)^{1/2}. \eqno(2e)$$
$a$ is the QCD axion and $a^\prime$ is the $E_8^\prime$ axion.
$a^\prime$ is massless at high temperatures.  But it acquires a
potential and a mass $m$ as the universe cools.  The
low temperature $(T<\Lambda)$ potential for
$a^\prime$ is calculated to be\refto{choikim85b,dineroseiwit}
$$V={\pi\over 4 M^2_{Pl}} \Lambda^6
\big|1-\rm{exp(i}a^\prime/15 F_a^\prime)\big|^2.
\eqno(3)$$
Therefore the low temperature mass of the axion $a^\prime$ is
$$m_0^2={\pi\over 450}{\Lambda^6\over M^2_{Pl}F_a^{\prime 2}}. \eqno(4)$$

$a^\prime$
decays to two gluons, photons or $Z'{\rm s}$ or to a $W^\pm$ pair
through the coupling $a^\prime F\tilde {F}$.  The gluons, $W^\pm$ and $Z$
particles predominantly create jets of mesons and some baryons.
At $T\sim \Lambda$, the non-abelian subgroup in the hidden sector gets strong
leading to the hadronization of particles.  The hidden sector hadrons will
have masses of the order of $\Lambda$.  Therefore we can ignore decays to
hidden sector particles.
A large fraction of the energy released as mesons
is transferred
to the photons, either through scattering or annihilations.  We shall assume
that
only 20 per cent of the energy of the decaying axions is transferred to the
photons i.e. $f=0.2$.  This is an extremely conservative estimate.

The lifetime of the axion is given by
$$\tau={8200 \pi^5} (g_1^4 +g_2^4 +g_3^4)^{-1} {F_a^{\prime 2}\over m_0^3}
=5.2 \times 10^{-17} {\rm{s}}
(g_1^4 +g_2^4 +g_3^4)^{-1}{\pi^{7/2}}{F_a^{\prime 5} M_{Pl}^3 \gev\over
 \Lambda^9} \eqno(5)$$
where $g_{1,2,3}$ are the couplings of $U(1)_Y$, $SU(2)_L$ and $SU(3)$
respectively.  We have set $F^2_2-F_1^2 \over F_1^2+F_2^2$ equal to 1.
We have checked that induced decays to photons and gluons due to
coherent axion field oscillations\r{pww83,as83,resaxdecay,koflinstar} are
inhibited due to
Hubble expansion.
For $\tau$ between $1$ s and about $10^{13}$ s,
the axion mass ranges between
$10^4 \gev$ and about $10^{-1} \gev$ and the axion decays
predominantly to mesons
through gluons.  For this range of lifetimes
one can
ignore $g_1$ and $g_2$ and let $g_3 =1$ (or $\alpha_s\sim 0.1$).
For lifetimes greater than about $10^{13}$ s the axion mass is less than
twice the pion
mass and the axion can only decay to photons.  If one rewrites eqn. (2a) in
terms of $F_{EM}{\tilde F}_{EM}$, one gets the lifetime by replacing
$( g_1^4 +g_2^4 +g_3^4)^{-1}$ in eqn. (5) by $(4e^4)^{-1}$.

{\centerline{\bf III}}
When the Peccei-Quinn symmetry breaks at early times the
axion field need not be at the minimum of its potential.   At a temperature
$T_i$ the field will
start to oscillate about the minimum of its potential with a period
$m^{-1}(T)$.
(Above $T_i$ the period of the oscillations is greater than the age of the
universe $~H^{-1}$.  At high
temperatures the mass has a temperature dependence, just as
for the QCD axion.)
We study the zero momentum mode of the $E^\prime_8$
axion field which can be treated
as a condensate
of zero momentum particles.
(Higher momentum modes are
redshifted.)
The energy density of the axion field today is given by\r{pww83,as83,df83}
$$\rho_{a_0}=m_0n_0={1\over2}m_0^2 A_0^2 \eqno(6)$$
$n_0$ is the number density today and $A_0$ is the amplitude of the
oscillations
today.
A classical treatment is permitted as the number density is
large\r{lindebk}.
The energy density can be rewritten in terms of parameters at $T_i$ when
the field starts oscillating as
$$\rho_{a_0}=m_0 n_i \Bigl ( {g_{s_0}\over g_{s_i}} \Bigr) {T_0^3\over
T_i^3}
      ={1\over2} m_0 m_i A_i^2
\Bigl ({ g_{s_0}\over g_{s_i}} \Bigr) {T_0^3\over T_i^3}. \eqno(7)$$
$g_{s}$ is the effective number of relativistic degrees of freedom
used to calculate the entropy.
The above expression assumes that the axion field loses energy only due to the
expansion of the universe and the variation of the mass with temperature and
not due to decays.
(The astrophysical constaints on massive decaying particles is given in
terms of limits on the ratio of the energy density of the massive
particles today, if they had not decayed, to the photon number density
today.)
$T_i$ is determined by the condition $m_i=3H$\r{pww83}.
$$T_i=\Biggl[ { m_i M_{Pl} \over 5 g_{*i}^{1/2} }\Biggr]^{1/2} \eqno(8)$$
$g_{*}$ is the effective number of relativistic degrees of freedom
used to calculate the energy density.
We shall take $g_{s_i}=g_{*i}$ and $g_{s_f}=3.9$.
By the time the axion field starts oscillating, the axion mass has
already attained its low temperature value in (4).  One can verify this
by checking that $m<3H$ at $T=\Lambda$ when the field attains its low
temperature mass (Appendix B).
In deriving (8) we have assumed that the universe
is radiation dominated when the axion field starts to oscillate.  One can
verify this by checking that $m>3H$ at $T_{eq}$ when
$\rho_a=\rho_r$, i.e., the axion field starts oscillating before the axion
energy density dominates the universe (Appendix C).
Combining (7) and (8) gives
$$\rho_{a_0}=
{5^{3/2}\over2}{g_{s_0}\over g_{*i}^{1/4}} m_0^{1/2} A_i^2
{T_0^3\over M_{Pl}^{3/2}}. \eqno(9)$$
$A_i$ is taken to be $F_a^\prime\simeq 10^{15}\gev$.
In a non-inflationary universe, the spatial average of $A_i/F_a^\prime$ is
$\pi/\sqrt 3$.  In an inflationary universe, setting $A_i$ equal to
$F_a^\prime$ assumes that the universe lies in a typical post-inflationary
region with $A_i/F_a^\prime \sim 1$.  Linde has criticized this
assumption.  We refer the reader to Ref.\r{lindeaxion} for a discussion
of this issue.
Since the number density of photons $n_\gamma=(1.2/\pi^2)g_\gamma T^3$
$$\rho_a/n_\gamma\Big |_{0}=0.62 \pi^{9/4}
{\Lambda^{3/2} F^{\prime 3/2}_a \over M_{Pl}^2} . \eqno(10)$$
This is the ratio of the axion energy density today in the absence of decays
to the  photon number density today.

Our method will be to look at the astrophysical and cosmological constraints
that apply to $\rho_a/n_\gamma\Big |_ {0}$ for different
lifetimes $\tau$.
For lifetimes between 1 s and $1.1\times 10^7$ s
we calculate the range of $\Lambda$ from
(5) and substitute that in (10) and compare with the limits on the energy
density
from
observations.
We show that axions with lifetimes
in this range
are ruled out
by cosmological arguments for $F_a^\prime=10^{15}\gev$.
Larger lifetimes are ruled out by limits on the distortions of
the microwave and gamma ray backgrounds and by closure arguments.
They would also give too small masses for the
gauginos and supersymmetric scalars.
$\tau$ less than 1 s corresponds to $\Lambda$ greater than
$1.2\times10^{13}\gev$.

{\centerline{\bf IV}}

The presence of the axion has different effects during different epochs in the
evolution of the universe.  The presence of axions during the neutron-proton
freeze-out ($\sim 1$ s) affects the ratio of neutrons to protons at freeze-out
and thereby leads to a change in the amount of helium in the universe.
Their presence during light element synthesis ($\sim 1$ minute) affects the
amount of deuterium and $\rm{^3He}$.  The decay of axions to photons
or to other particles that transfer their energy to photons increases
$n_\gamma$ and dilutes $\eta$, the ratio of the baryon number density to
the photon number density.  This then alters the calculations of light element
abundances.
For $10^4$ s $<\tau<10^7$ s, the
decay photons interact with the background thermal photons and create a
cascade of photons with enough energy to photodissociate deuterium and
helium.  For lifetimes between $10^5$ s and (re)combination
the decay photons change the shape of the
spectrum of the cosmic background radiation.  Decays occuring after
recombination
change the diffuse extra-galactic gamma ray spectrum.  Decays of lifetimes
greater than the age of the universe can lead to the axion energy density
overclosing the universe.

We now explore the astrophysical constraints on the energy density of the
axions for different lifetimes of the axion.

{\bf 1 s ${\bf <{\bf \tau}< 10^4}$ s:} Increased expansion rate and dilution of
$\eta$

The interaction rate for the weak reactions
$n+\nu_e\leftrightarrow p+e^-$ and $p+\bar \nu_e \leftrightarrow n+e^+$
goes as $G_F^2 T^5$ while the expansion rate of the universe is
$\rm{H}=[(8\pi/3) \rho/M_{Pl}^2]^{(1/2)}$, where $\rho$ is the energy density
of
the universe\r{kolbt}.
When the weak reactions are faster than the expansion of the universe
the ratio of the neutrons to protons is given by
$${n\over p}={\rm {exp}} \Big [-{\Delta m \over T}\Bigr]. \eqno(11)$$
$\Delta m$ is the difference between the neutron and the proton mass.
In the standard cosmology, at $t\sim 1$ s $(T\sim 1 \mev)$, the reaction rates
become
less than the expansion rate of
the universe and the neutrinos freeze out and
the protons and neutrons are no longer in chemical
equilibrium.  Thereafter the neutron-proton ratio does not change from its
equilibrium value at the freeze-out temperature, except for free neutron
decays.  The precise temperature at which the freeze-out occurs is determined
by the expansion rate of the universe which is influenced by the energy density
of the axion.  At slightly later times between 1 and 3
minutes $(0.3\mev >T>0.1 \mev)$, the light elements deuterium,
$\rm {^3He}$, $\rm {^4He}$ and
$\rm {^7Li}$ are produced.

An increase in the expansion rate due to the axions causes the weak reactions
above to freeze out at an earlier temperature.  This leads to a
a higher neutron to proton ratio and a higher percentage
of $\rm {^4He}$, if all the neutrons are converted into helium.
However, the deuterium and $\rm{^3He}$ abundances are
sensitive to the expansion rate
at temperatures between $0.08 \mev$ and
$0.04 \mev$ when the strong reactions that convert deuterium
and $\rm{^3He}$ into $\rm{^4He}$
freeze out.
The amount of deuterium, $\rm {^3He}$ and $\rm {^4He}$ produced
also depends on $\eta_{ns}$, the baryon-to-photon ratio during
nucleosynthesis\r{schertur88a}.
For a given $\eta_0$ today, the baryon-to-photon ratio before the axions
decay is higher for higher
axion energy density.  The amount of $\rm {^4He}$ today increases for
higher $\eta_{ns}$ while
the amount of deuterium and $\rm {^3He}$ decreases because the rates of the
reactions that convert deuterium and $\rm {^3He}$ into
$\rm {^4He}$ are
 proportional
to $\eta_{ns}$.

The constraint that  $\rm {^4He}$
makes up less than $25$ percent of the energy density of the baryons in the
universe and that
the ratio of the abundances of deuterium and  $\rm {^3He}$ to the
baryonic abundance is less than $1.0\times 10^{-4}$
gives us limits on $\rho_a/n_\gamma \Big|_{0}$.  For 1 s $ <\tau< 10^4 $ s,
the limit from light element abundances implies that $\rho_a/n_\gamma
\Big |_{0}$
is less than
$1.5\times 10^3 f^{-1}\gev$
to $8.4\times 10^{-5} f^{-1}\gev$\r{schertur88a}.
But the axion energy density lies
between $7.5\times 10^4\gev$ and
$1.6\times 10^4 \gev$.
Therefore this
range of lifetimes corresponding to $\Lambda$ between $1.2\times 10^{13}\gev$
and
$4.3\times 10^{12}\gev$ and the axion mass between $1.2\times 10^4 \gev$
and $5.5\times 10^2 \gev$  is ruled out.

{\bf ${\bf 10^{4}}$ s ${\bf <\tau<1.1\times 10^{7}}$ s:}
Photodissociation

For decays that occur in the time interval
$10^{4}$ s $<\tau<10^{7}$ s,
the
decay photons can initiate electromagnetic cascades that can photodissociate
deuterium and helium\r{lindley}.
The high energy decay photons interact with the thermal
photons producing electron-positron pairs.  The electrons and positrons then
lose their energy by inverse-Compton scattering the thermal photons to higher
energies.  These photons will repeat the above process till their energies
fall below $E_* \simeq m_e^2/22T$, the threshhold for pair production.
($m_e$ is the electron mass.)
Photons with energy greater than $E_*$ and $Q$, the threshhold
for photodissociation of an element, are more likely to pair produce than
interact with the nuclei of the light elements\r{lindley}.
The spectrum of the electromagnetic cascade
initiated by a decay photon of energy $E ( > E^*)$ is\r{ellispoulossark92}
$${dN^E \over dE_\gamma} =\cases{
{24 \over 55}\sqrt {2} {E\over \sqrt {E_*}} E_\gamma^{-3/2}
&for $0\leq E_\gamma \leq E_*/2$ \cr
{3\over 55} E E_*^3 E_\gamma^{-5}
&for $E_*/2 \leq E_\gamma \leq E_*$ \cr
0 &for $ E_* <  E_\gamma $,\cr} \eqno(12)$$
where the cascade spectrum has been normalized as
$${\int_0^{E_*} E_\gamma \Bigl ( {dN^E \over dE_\gamma} \Bigr ) \, dE_\gamma}
=E.  \eqno(13)$$
The photodissociation of an element starts when $E_*$ becomes larger than
the photodissociation threshhold for that element.

The mass fraction of an element is defined as
$X_k\equiv n_k Z_k/n_N$, where $Z_k$ is the atomic number of the element,
$n_k$ is the number density of the element
and  $n_N$ is the number density of the nucleons.  The change in
the mass fraction is given by\refto{lindley,ellispoulossark92}
$$
\eqalignno{
\quad\quad\quad\quad\quad
{dX_k \over dt}=-{dn_a \over dt}
\int_0^\infty {dN\over dE}\,dE
&\Bigl[
\int^{E_*(t)}_{Q_k}{dN^E \over dE_\gamma}
{X_k \sigma_k \over n_e \sigma_C} \, dE_\gamma
\cr
&-\sum_{j \ne k}
{\int^{E_*(t)}_{Q_j} {dN^E \over dE_\gamma}
{X_j \sigma_{j \rightarrow k} \over n_e \sigma_C} \,
dE_\gamma}
\Bigr ]
\quad\quad\quad\quad\quad
\quad(14)
\cr
}$$
where $\sigma_{j\rightarrow k}$ is the partial
cross-section for the dissociation of element $j$ to $k$, $\sigma_k$ is the
total cross-section for photodissociation of element $k$ and $\sigma_C$ is the
Compton scattering cross-section on the thermal electrons with density $n_e$.
Anticipating the stringent conditions on the energy density of axions
we assume above that
$n_e$ is not changed significantly by the decays and cascade process.
${dn_a \over dt}$ includes the change in the number density of axion due to
decays
only (and not due to the expansion of the universe).
$${dn_a \over dt}=-{n_a^I(t)\over \tau}\rm{exp} \Bigl(-{t \over\tau} \Bigr)
 \eqno(15)$$
where $n_a^I(t)$ is the value of $n_a$ at some very early time $t_I$
much less
than $\tau$, scaled to the time of dissociation $t$.  $e^{-t_I/\tau} \sim 1$.
${dN\over dE}$ is the spectrum of decay photons.
To simplify our calculation we only include photons coming directly
from the decay of the axions.  ${dN\over dE}$ is then
$2\delta(E-m_0/2)$.  The ratio of the decay rates into photons and into gluons
is $\alpha_{EM}^2/\alpha_s^2 \sim 10^{-5}$.  Therefore we include a factor
of $f_\gamma=10^{-5}$ below.  Including all the energy transferred to photons
via all the axion decay channels can only tighten our constraints.

Integrating (14) gives
$$X^f_k-X^i_k \simeq f_\gamma {\rho_a\over n_\gamma} \Bigg |_{0}
         {8\over7 \eta_{0}}
                \big[-X^i_k\beta_k(\tau)+
         \sum_{j \neq k}X_j^i\beta_{j\rightarrow k}(\tau) \big]  \eqno(16a)$$
where
$$\beta_k(\tau)={1\over \tau}
{\int^\infty_{t_k^i} {\rm dt} \, {\rm exp}\Bigl(-{{\rm t} \over\tau} \Bigr)
{\int^{E_*(t)}_{Q_k} {1\over E} {dN^E\over dE_\gamma} {\sigma_k(E_\gamma)\over
\sigma_C(E_\gamma)}\, dE_\gamma}} \eqno(16b)$$
and
$$\beta_{j\rightarrow k}(\tau)={1\over \tau}
{\int^\infty_{t_j^i} {\rm dt} \, {\rm{exp}} \Bigl(-{t \over\tau} \Bigr)
{\int^{E_*(t)}_{Q_j} {1\over E}
{dN^E\over dE_\gamma} {\sigma_{j\rightarrow k}(E_\gamma)\over
\sigma_C(E_\gamma)}\, dE_\gamma}}
. \eqno(16c)$$
$i$ above represents the time when the photodissociation starts,
i.e., when the cascade cutoff energy $E_*$ equals the photodissociation
threshold $Q$.
$\eta_0$ is the baryon-to-photon ratio today.
Above we have used
$$n_a^I(t)/n_e(t)=n_a^I(t_0)/n_e(t_0) \eqno(17)$$
and
$$n_e(t_0) \simeq {7\over 8} n_{N_0}={7\over 8}\eta_0 n_{\gamma_0}.
\eqno(18)$$
The subscript $0$ refers to quantities today and $N$ to nucleons.

$t^i$ is
calculated from
$$t^i=1.6\times10^{-25}{\rm s}  g_{s}^{-1/2}
{M_{Pl}^{7/4} \gev\over m_0^{1/4} F_a^\prime (m_e^2/22Q)^{3/2}}
\eqno(19)$$
where $Q$ is $2.2\mev$ for deuterium,
$6.5 \mev$ for $\rm {^3He}$ and $20 \mev$ for $\rm
{^4He}$\r{ellispoulossark85}.
We have used the time-temperature relationship obtained in eqn.(11) of
ref.\r{ranga94a} to obtain (19).
$g_{s}$ is 3.9.
The photodissociation of
light elements
only goes on till the energy of
the decay photons itself falls below the threshhold for pair production.
Then no cascade appears and photodissociation effects can be ignored.
The time at which this occurs is calculated from
$$E={m_0\over 2}={m_e^2\over 22T} \eqno(20a)$$
where
$$m_0=1.2 \times 10^{-6} \Bigl[ {F_a^{\prime 2} \gev
\over (\tau/\rm{s})}\Bigr]^{1/3}
\eqno(20b)$$
(20b) is calculated from (5).  Combining (20) with the $t-T$ relationship
$$T=\Big[ {{\rm s}\over t} {1.6\times 10^{-25} \over g_{s}^{1/2}}
{M_{Pl}^{7/4} \gev \over m_0^{1/4} F_a^{\prime}}  \Big ]^{2/3} \eqno(21)$$
gives $t=1.1\times 10^{7}$ s.
For decays that occur after this time, photodissociation of
light elements
is negligible.

For $10^{4}$ s $<\tau<1.1\times 10^{7}$ s,
the limits on $\rho_a/n_\gamma\Big |_0$
come from the constraints on the increase in deuterium and $\rm{^3He}$
due to photodissociation
of $\rm {^4He}$.  (The depletion of deuterium and $\rm{^3He}$ is
eclipsed by the increase in deuterium and $\rm{^3He}$ due to
photodissociation
of $\rm {^4He}$ as $\rm {^4He}$ is $10^4$ times more abundant.)
$$X^f(D+{\rm ^3He}) -X^i(D +{\rm{^3He}})\simeq {8\over 7} Y^i({\rm {^4He}})
{\beta_{\rm {^4He}}(\tau)\over \eta_{0}} r
{\rho_a \over n_\gamma} \Bigg |_{0} f_\gamma,\eqno(22)
$$
where $r=
({3\over4}\sigma_{\gamma {\rm ^4He}\rightarrow n {\rm ^3He},p {\rm ^3H}}
+{1\over2}\sigma_{\gamma {\rm ^4He}\rightarrow n p D})/
\sigma_{\gamma {\rm ^4He}\rightarrow all}\simeq 0.5$.

$\beta$ are computed numerically in Ref.\r{ellispoulossark92}
for a radiation dominated universe.  The $t^i$ in our matter dominated
universe are much earlier than those calculated in
Ref.\r{ellispoulossark92}.
Hence $\beta_{\rm {^4He}}$ will be greater in our case for lifetimes
between $10^4$ s and $10^7$ s.  We take $\beta_{\rm {^4He}}$ to be the
peak value obtained in Ref.\r{ellispoulossark92} of $1\gev^{-1}$.
We take $Y^i(\rm{^4He})<.25$ which implies $\eta< 1.3\times 10^{-9}$ and
consequently $X^i(D+\rm{^3He})> 3.4 \times 10^{-5}$.
To be consistent with galactic chemical evolution $(D+\rm{^3He})/H
<10^{-4}$
or $X^f(D+\rm{^3He})<2.3\times 10^{-4}$.  Thus the upper limit on
${\rho_a /n_\gamma}\Big |_{0}$ is $1.0\times10^{-12}f^{-1}_\gamma\gev$.
But the calculated value of
${\rho_a /n_\gamma}\Big |_{0}$ is between $1.6\times 10^4\gev$ and
$5.1\times10^3\gev$.
Hence the corresponding values of $\Lambda$ between
$4.3\times10^{12}\gev$
and $2.0\times 10^{12}\gev$ are ruled out.  The corresponding values of
the
axion mass lie between $5.5\times 10^2 \gev$ and $56 \gev$.

{\bf ${\bf \tau>1.1 \times 10^7}$ s:}

For lifetimes greater than $1.1 \times 10^7$ s, the decay photons would
distort the cosmic microwave background or show up in the gamma ray
background spectrum.  Such lifetimes would also
gives supersymmetric masses less than $100 \mev$.  Lifetimes greater than
the age of the universe would overclose the universe.
Furthermore, if one assumes that the maximum baryon asymmetry
$\eta\equiv n_B/n_\gamma$ produced in
the early universe is O(1)\r{affledine,linde} and that the baryon asymmetry
today is $3\times 10^{-10}$, then constraints on
the dilution of the universe imply that $\Lambda$ must be greater than
$7\times 10^{12}\gev$\r{ranga94a}.

{\bf Conclusion:}

Thus we see that astrophysical and cosmological arguments imply that
$\Lambda$ is greater than $1.2\times10^{13}\gev$ or that $M_S$ is
greater than $10^{10}\gev$.  Our result agrees with
the value of $5\times10^{13}\gev$ for $\Lambda$ obtained independently by
setting supersymmetric scalar masses
equal to $m_W$.

I would like to thank Mark Srednicki, Subir Sarkar, Robert Scherrer and
Kiwoon Choi for very useful
discussions.  I would also like to thank the referee for pointing out
the possibility of the induced decay of axions due to resonance
effects.
I am grateful to the Center for Particle Astrophysics
at the University of California, Berkeley, where
most of this work was completed, for their hospitality.

This work was supported by NSF Grant No. PHY91-16964.

\endpage

\centerline{\bf Appendix A}

In this Appendix, we show that the universe is radiation dominated at
the
temperature $T\sim \L$, when the axion attains it low temperature mass.

At $T=\L$,  the radiation energy density is given by
$$\rho_{rad}={\pi^2\over 30} g_* \L^4 \eqno(A.1)$$
while the axion energy density is given by
$$\eqalignno{\rho_a
=m_0 n
&={1\over2}m_0^2 A^2(\L)\cr
&={\pi\over 900}\L^4 {\L^2\over M_{Pl}^2}{A^2(\L)\over F_a^{\prime 2}}
&(A.2)\cr}$$
The low temperature mass $m_0$ used above is given by
(4).
$g_*$ is 106.75. At $T=\L$ the field has not started oscillating (see
Appendix B) and
we assume that $A(\L)$ is approximately $F_a^\prime$.
Also, $\L<M_{Pl}$.  Thus, $\rho_a<\rho_{rad}$ at $T=\L$.

\centerline{\bf Appendix B}

In this Appendix we show that the axion field starts oscillating at
a temperature
$T_i$
below $T=\L$ at which the axion attains
its low temperature mass.  We shall show below that $m<3H$ at $T=\L$,
i.e., the axion field has not yet started oscillating at $T=\L$.

Since the universe is still radiation dominated
at $T=\L$ (see Appendix A),
$$3H=0.9 \pi^{3/2} g_*^{1/2} {\L^2\over M_{Pl}}
\eqno(B.1)$$
However,
$$m_0=\Biggl ({\pi\over 450}\Biggr)^{1/2} {\L^2\over M_{Pl}}{\L\over
F_a^\prime}
\eqno(B.2)$$
Since $\L<F_a^\prime$ and $g_*=106.75$, $m<3H$ at $T=\L$.
Thus the axion field starts oscillating at a temperature
$T_{i}<\L$,
when $H$ has decreased sufficiently to satisfy $m=3H$.

\endpage

\centerline{\bf Appendix C}

In this Appendix we show that the universe is still radiation dominated
at the temperature $T_i$
when the axion field starts
oscillating.
We actually show that the axion field is already oscillating by the time
the
temperature has decreased to $T_{\rm eq}$, when the axion energy density
is
equal to the radiation energy density.

At $T=T_{\rm eq}$,
$$3H=(12\pi)^{1/2} {A(T_{\rm eq})\over M_{Pl}} m_0 \eqno(C.1)$$
Since $A(T_{\rm eq})\leq A(T_i)=F_a^\prime<0.2 M_{Pl}$,
$m>3H$ at $T_{\rm eq}$.  This implies that the universe started
oscillating
at a temperature greater than $T_{\rm eq}$, when the universe was still
radiation dominated.

\references

\hyphenation{Amsterdam}

\refis{ranga94a}R. Rangarajan, \np, B454, 1995, 369.

\refis{r6}R. Rangarajan, \np, B454, 1995, 357.

%

\refis{pww83}J. Preskill, M. B. Wise, and F. Wilczek,
\pl, B120, 1983, 127.

\refis{as83}L. Abbott and P. Sikivie, \pl, B120, 1983, 133.

\refis{df83}M. Dine and W. Fischler, \pl, B120, 1983, 137.

\refis{lindeaxion}A. D. Linde, \pl, B201, 1988, 437; \pl, B259, 1991, 38.

\refis{resaxdecay}I. Tkachev, {Sov. Astron. Lett.} {12} (1986) 305.

\refis{koflinstar}L. A. Kofman, A. D. Linde and A. A. Starobinsky, hep-th
9405187 (1994).

\refis{srednicki90}Particle Physics and Cosmology: Dark Matter,
ed. M. Srednicki (North--Holland, Amsterdam, 1990).

\refis{turner86}M. S. Turner, \prd, 33, 1986, 889.  

\refis{witten85}E. Witten, \pl, B155, 1985, 151. 

\refis{linde83}A. D. Linde, \pl, 129B, 1983, 177.

\refis{ghmr}D. J. Gross, J. A. Harvey, E. Martinec and R. Rohm,
\prl, 54, 1985, 502.

\refis{greenschwartz}M. B. Green and J. H. Schwarz, \pl, B149,
1984, 117.

\refis{ghmrgreenschwartz}D. J. Gross, J. A. Harvey, E. Martinec and R. Rohm,
{\it Phys. Rev. Lett.} {\bf 54}, (1985), 502;
M. B. Green and J. H. Schwarz,
{\it Phys. Lett.} {\bf B149}, (1984), 117.

\refis{choikim85a}K. Choi and J. E. Kim, \pl, B154, 1985, 393.

\refis{choikim85b}K. Choi and J. E. Kim, \pl, B165, 1985, 71.

\refis{kim87}J. E. Kim, \prep, 150, 1987, 1.

\refis{sred85}M. Srednicki,
\np, B260, 1985, 689.

\refis{kap85}D. Kaplan,
\np, B260, 1985, 215.

\refis{geokapran86}H. Georgi, D. Kaplan and L. Randall,
\pl, B169, 1986, 73.

\refis{deribnil}J. P. Derendinger, L. E. Ibanez and H. P. Nilles,
\pl, B155, 1985, 65.

\refis{dineroseiwit}M.Dine, R. Rohm, N. Seiberg and E. Witten, \pl,
B156, 1985, 55.

\refis{barrchoikim}S. M. Barr, K. Choi, and J. E. Kim, \np, B283, 1987,
591.

\refis{barr85}S. M. Barr, \pl, B158, 1985, 397.

\refis{wenwit}X. G. Wen and E. Witten, \pl, B166, 1986, 397.

\refis{dsww}M. Dine, N. Seiberg, X. G. Wen and E. Witten, \np, B289,
1987, 319.

\refis{wenwitdsww}X. G. Wen and E. Witten,
{\it Phys. Lett.} {\bf B166} (1986) 397;
M. Dine, N. Seiberg, X. G. Wen and E. Witten, {\it Nucl. Phys.}
{\bf B289} (1987) 319.

\refis{witten84}E. Witten, \pl, B149, 1984, 351.

\refis{witten85}E. Witten, \pl, B153, 1985, 243.

\refis{witten85b}E. Witten, \pl, B155, 1985, 151.

\refis{canhorstrowit}P. Candelas, G. T. Horowitz, A. Strominger and
E. Witten, \np, B258, 1985, 46.

\refis{lindebk}A. D. Linde, Particle Physics and Inflationary Cosmology,
(Harwood Academic Publishers, Switzerland, 1990).

\refis{pwwasdf83}J. Preskill, M. B. Wise, and F. Wilczek,
{\it Phys. Lett.}, {\bf B120} (1983) 127;
L. Abbott and P. Sikivie, {\it ibid.}, 134;
M. Dine and W. Fischler, {\it ibid.}, 137.


\refis{schertur88a}R. J. Scherrer and M. S. Turner, \apj, 331, 1988, 19.
The limits calculated in this paper are on $\rho/n_\gamma$ at $T=10^{12} K$.
To obtain limits on $\rho_a/n_\gamma \Big |_0$ we multiply their results by a
factor of $4/11$ to account for $e^\pm$ annihilation.

\refis{schertur88b}R. J. Scherrer and M. S. Turner, \apj, 331, 1988, 33.

\refis{wagoner}R. V. Wagoner, \apj, 179, 1973, 343.

\refis{ellispoulossark85}J. Ellis, D. V. Nanopoulos and S. Sarkar, \np,
B259, 1985, 175.

\refis{ellispoulossark92}J. Ellis, G. B. Gelmini, J. L. Lopez,
D. V. Nanopoulos and S. Sarkar, \np,
B373, 1992, 399.  This is a comprehensive paper on astrophysical
constraints on massive decaying particles.

\refis{sarkcoop}S. Sarkar and A. M. Cooper, \pl, B148, 1984, 347

\refis{kolbscherr}E. W. Kolb and R. J. Scherrer, \prd, 25, 1982, 25.

\refis{kolbt}E. W. Kolb and M. S. Turner, The Early Universe
(Addison-Wesley Publishing Company, 1990).

\refis{lindley}D. Lindley, \apj, 294, 1985, 1.

\refis{husilk}W. Hu and J. Silk, \prl, 70, 1993, 2661.

\refis{mather}J. C. Mather et al, \apj, 420, 1994, 439.

\refis{danzo}L. Danese and G. De Zotti, \journal Riv. Nuovo Cimento Soc.
Ital. Fis., 7, 1977, 277.

\refis{daly}R. Daly, \apj, 324, 1988, L47.

\refis{zds}A. Zdziarski and R. Svensson, \apj, 344, 1989, 551.

\refis{schgrp}V. Schonfelder, F. Graml and F. P. Penningsfeld, \apj,
240, 1980, 350.

\refis{schgd}V. Schonfelder, U. Graser and J. Daugherty, \apj,
217, 1977, 306.

\refis{trombka}J. Trombka, C. S. Dyer, L. G. Evans,
M. J. Bielefield, S. M. Seltzer, A. E. Metzger, \apj, 212, 1977, 925.

\refis{fichtel}C. E. Fichtel, R. C. Hartman, D. A. Kniffen,
D. J. Thompson, H. B. Ogelman and M. E. Ozel, \apj, 222, 1978, 833.

\refis{trombfich}J. Trombka and C. E. Fichtel, \prep, 97, 1983, 173.

\refis{olivsilk}K. Olive and J. Silk, \prl, 55, 1985, 2362.

\refis{schturn85}R. J. Scherrer and M. S. Turner, \prd, 31, 1985, 681.

\refis{affledine}I. Affleck and M. Dine, \np, B249, 1985, 361.

\refis{linde}A. D. Linde, \pl, B160, 1985, 243.

\refis{affledinelinde}I. Affleck and M. Dine, {\it Nucl. Phys.}
{\bf B249}, (1985)
361; A. D. Linde, {\it Phys. Lett.} {\bf B160}, (1985) 243.

\endreferences
\end